\documentclass[10pt,letterpaper]{article}
\usepackage[left=1in,right=1in,top=1in,bottom=1in]{geometry}
\usepackage{amsmath,amssymb}
\usepackage{graphicx}
\usepackage{float}
\usepackage[labelfont=bf]{caption}
\usepackage{titlesec}
\usepackage[colorlinks=true,allcolors=blue]{hyperref}
\usepackage{enumitem}
\usepackage{subcaption}
\usepackage{tikz}
\usepackage{authblk}
\usepackage{mathrsfs}

\titlespacing*{\section}{0pt}{1.2em}{0.6em}
\titlespacing*{\subsection}{0pt}{1.0em}{0.4em}

\DeclareMathOperator{\erf}{Erf}
\DeclareMathOperator{\erfc}{Erfc}

\title{Extracting Spectral Diffusion in Two-Dimensional Coherent Spectra via the Projection Slice Theorem}

\author[1]{Cesar Perez} 
\author[1, 2]{Steven Cundiff\thanks{cundiff@umich.edu}}

\affil[1]{Department of Physics, University of Michigan, Ann Arbor, MI 48109, USA}
\affil[2]{Quantum Research Institute, University of Michigan, Ann Arbor, MI 48109, USA}

\date{\today}

\begin{document}

\maketitle

\begin{abstract} 
A robust and streamlined method is presented for efficiently extracting spectral diffusion from two-dimensional coherent spectra by employing the projection-slice theorem. The method is based on the optical Bloch equations for a single resonance that include a Frequency-Frequency Correlation Function (FFCF) in the time domain. Through the projection slice theorem (PST), analytical formulation of the diagonal and cross-diagonal projections of time-domain two-dimensional spectra are calculated that include the FFCF for arbitrary inhomogeneity. The time-domain projections are Fourier transformed to provide frequency domain slices that can be fit to slices of experimental spectra. Experimental data is used to validate our lineshape analysis and confirm the need for the inclusion of the FFCF for quantum wells that experience spectral diffusion. 

\end{abstract}

\section{Introduction}

Two-dimensional coherent spectroscopy (2DCS) is a powerful nonlinear optical technique that reveals the complex dynamics of material systems by correlating their response to a sequence of ultrashort laser pulses \cite{2011Hamm_Book,2023Li_Book}. By spreading spectral information across two frequency dimensions, 2DCS resolves features often obscured in linear spectroscopy, such as the coupling between resonances, many-body interactions, and the distinct contributions of homogeneous and inhomogeneous broadening to a spectral lineshape \cite{2023Li_Book}. This capability has made it an indispensable tool for studying a vast range of systems, from atomic vapors \cite{2003Tian_Science,2010Dai_PRA,2012Dai_PRL,2013Li_NCommun} and molecules \cite{2000Asplund_PNAS,2001Golonzka_PRL,2011Hamm_Book} to semiconductor nanostructures \cite{2007Zhang_PNAS,2006Li_PRL,2021Purz_PRB}, and color centers in diamond \cite{2021Bates_JAP,2022Day_PRL,2021Smallwood_PRL,2021Liu_MQT}. A particularly elegant way to separate broadening mechanisms is through the projection-slice theorem (PST), which connects projections in the time domain to slices in the frequency domain \cite{1956Bracewell_AustJPhys}.

Beyond static broadening, 2DCS is exceptionally suited for tracking time-dependent energy fluctuations, a phenomenon known as spectral diffusion \cite{1995Mukamel_Book}. These dynamics, arising from the coupling between a system and its environment, are typically quantified by the frequency-frequency correlation function (FFCF) \cite{1995Mukamel_Book,2016Singh_JOSAB}, defined as

\begin{equation}
    C(T) = \frac{<\delta \omega(t-T) \delta \omega(t)>}{<\delta \omega^2>}
\end{equation}{}
where $\delta\omega(t) = \omega(t) - <\omega>$ is the time-dependent frequency offset from the mean value and $T$ is time.
{The FFCF formalism quantifies the magnitude of frequency correlation loss and is valid under the strong redistribution approximation, which assumes that energy fluctuations are symmetric.}

Experimentally, the FFCF is measured by acquiring 2D spectra as a function of the waiting time, T, and analyzing changes in the 2D peak shape \cite{2023Li_Book}. However, current methods for extracting the FFCF—such as analyzing the center-line slope (CLS) \cite{2015Sanda_JCPA,2016Singh_JOSAB}, peak ellipticity \cite{2006Roberts_JCP,2016Singh_JOSAB}, or fitting entire 2D spectra—can be indirect \cite{2016Singh_JOSAB}, computationally intensive, or rely on approximations that are not always accurate. Critically, conventional PST-based lineshape analysis has not incorporated the FFCF \cite{2010Siemens_OpEx}, leaving a disconnect between this powerful theorem and the direct quantification of system-bath dynamics.

In this work, we bridge this gap by presenting a streamlined and robust method that integrates the FFCF directly into the PST framework. We derive formulations for projections of the time-domain 2D signal that include the FFCF for an arbitrary degree of inhomogeneous broadening. By numerically Fourier transforming these 1D projections, we generate model frequency-domain slices that can be directly fit to experimental data. This approach allows for the simultaneous extraction of the homogeneous dephasing rate ($\gamma$),  the inhomogeneous broadening ($\sigma$), and the FFCF value ($C(T)$) from simple diagonal and cross-diagonal slices. Our method is agnostic to the specific functional form of the FFCF's decay and avoids both the geometric approximations of other techniques and the high computational cost of fitting full 2D spectra. We demonstrate the power of this technique by applying it to experimental 2D spectra of excitons in a semiconductor quantum well sample. The method proves to be a straightforward and versatile tool, establishing an improved standard for accurately extracting spectral diffusion dynamics from 2D coherent spectra.

\section{Theoretical Framework and Method}
To derive diagonal and cross-diagonal slices in the frequency domain, the starting point is the temporal signal calculated using perturbation theory to solve the optical Bloch equations under the assumption of instantaneous (delta function in time) optical pulses \cite{1979Yajima_JPSJ}. The three-pulse excitation scheme mentioned above is used. Gaussian statistics and the rotating wave approximation are assumed. The calculated signal in the direction ($\mathbf{k}_s=-\mathbf{k}_{A}+\mathbf{k}_{B}+\mathbf{k}_{C}$) is
\begin{equation}
    S(\tau,t) = A_{0} e^{-i\Delta (\tau -t)} e^{-\gamma (t+\tau)} e^{-\frac{1}{2}\sigma^2 (\tau-t)^2} \Theta (t) \Theta (\tau)
    \label{eqn:2Dtime_t-tau}
\end{equation}
where $\Delta$ is the detuning between the excitation resonance and the laser, $\gamma$ is the homogeneous dephasing rate, $\sigma$ is the width of the inhomogeneous distribution, $A_0$ is the signal amplitude, and $\Theta(t)$ is the Heaviside step function. By using rotated time coordinates $\tau'=t-\tau$ and $t'=t+\tau$, the signal can be decomposed into two components
\begin{equation}
    S(\tau',t')=A_0 e^{-(\gamma t' + i\Delta \tau' + \frac{1}{2}\sigma^2 \tau'^{2})} \Theta (t' - \tau') \Theta (t' + \tau').
    \label{eqn:2Dtime-t'-tau'}
\end{equation}
The first component describes to a homogeneous decay observed along the photon echo time axis $t'$, while the second component along the anti-echo time axis $\tau'$ oscillates at the detuning frequency and has a decay determined by the inhomogeneous width $\sigma$.

Using the rotated coordinates of $t'$ and $\tau'$ separates the signal decays due to the homogeneous dephasing and the inhomogeneous broadening in the time domain for slices of the two-dimensional data along $t'=0$ and $\tau'=0$. However this separation does not occur for slices in the frequency domain. According to the PST, a slice in the frequency domain is a projection in the time domain. Slices along the frequencies corresponding to $t'$ and $\tau'$, $\omega_{t'}$ and $\omega_{\tau'}$, do not fully separate the homogeneous and inhomogeneous lineshapes due to the Heaviside step functions, which enforce causality in the time domain. Thus the lineshapes of the frequency domain slices depend on both $\gamma$ and $\sigma$
 \cite{2010Siemens_OpEx}.

To extend this analysis to include spectral diffusion, we extend the analysis above to include the FFCF, $C(T)$ \cite{2001Everitt_JCP}. As mentioned above, the effects of spectral diffusion are typically observed by measuring 2DCS as a function of the waiting time between the second and third pulses, $T$. We assume that the effect of spectral diffusion during time periods $t$ and $\tau$ are captured in the dephasing rate. Thus the dependence on $T$ must be included in the time domain signal  
\begin{equation}
        S(\tau,T,t) = A(T) e^{-i\Delta (\tau -t)} e^{-\gamma (t+\tau)} e^{-\frac{1}{2}\sigma^2 (\tau^2 + t^2 -2C(T)\tau t)} \Theta (t) \Theta (\tau)
        \label{eqn:2Dtime-t-tau-FFCF}
        \end{equation}
where $A(T)$ is the $T$ dependent amplitude of the signal. We make the same coordinate transformation as for deriving Eq.~\ref{eqn:2Dtime-t'-tau'} yielding
\begin{equation}
    S(\tau',T,t')= A(T) e^{\frac{1}{4} ([C(T)-1]t'^{2} \sigma^{2} - [C(T)+1]\tau'^{2} \sigma^{2} - 4\gamma t' - 4i\Delta \tau')} \Theta (t' - \tau') \Theta (t' + \tau').
    \label{eqn:2Dtime-t'-tau'-FFCF}
    \end{equation}
{The full derivation of this coordinate transformation is provided in the Supplement \cite{Supplement}.} To check this result, we set $C(T)=1$ we find Eq.~\ref{eqn:2Dtime-t-tau-FFCF} equal to Eq.~\ref{eqn:2Dtime_t-tau} and Eq.~\ref{eqn:2Dtime-t'-tau'-FFCF} equal to Eq.~\ref{eqn:2Dtime-t'-tau'}, where the homogeneous and inhomogeneous contributions are completely correlated. These equations now allows us to employ the PST and include FFCF is our line shape analysis and simulations.

We extend this analysis to the 2D frequency domain through the application of the PST. According to the PST, Fourier-transforming a projection in one domain is equivalent to a slice in the 2D Fourier pair domain, where a projection onto a line in a specified direction is achieved by integrating the signal perpendicular to that line at each point \cite{1956Bracewell_AustJPhys,1978Nagayama_JMR}. Thus, Fourier transforming the 2D time domain data projected onto a line at an angle $\theta$ with respect to the $t$ axis produces a slice in the 2D frequency domain, at the same angle $\theta$ from the $\omega_t$ axis. In a 2D frequency spectrum, a signal oscillating at frequency $\omega_0$ will shift along the $\omega_{\tau'}$ axis by $\omega_0$ allowing for precise energy selection. Notably, if the laser frequency deviates from the resonance frequency, the resulting oscillations null the signal during the projection operation, illustrating how 2D spectroscopy can discern resonances or variations within an inhomogeneous distribution. The separation of homogeneous and inhomogeneous broadening is most evident along the diagonal and cross-diagonal in 2D frequency space, corresponding to slices along the $\omega_{t'}$ and $\omega_{\tau'}$ directions \cite{2010Siemens_OpEx}. Consequently, we assess projections onto the $\tau'$ and $t'$ axes in the time domain by integrating the signal perpendicular to the axis, adjusting the limits of integration to accommodate time ordering constraints. This approach does not require the limit of strong inhomogeneous broadening but works for any ratio of the inhomogeneous to the homogeneous widths. The time domain signal projected on the $t'$ axis is
\begin{equation}
    \begin{gathered}
    S_{proj,\omega_0} (t',T) =  
    \Theta(t') A(T)
    e^{\frac{1}{4}([C(T)-1]t^{'2} \sigma^2 - 4\gamma t')} \int_{-t'}^{t'} e^{-\frac{1}{4} [C(T)+1]\tau^{'2} \sigma^2} d\tau'    \\
    =   \\  
    \frac{2\sqrt{\pi} \Theta(t') A(T) e^{\frac{1}{4}t'(-4\gamma + [C(T)-1]t'\sigma^2)}\erf{[\frac{1}{2}\sqrt{1+C(T)}t'\sigma]}}{\sqrt{1+C(T)}\sigma} 
    \end{gathered}
    \label{eqn:time-proj-t-FFCF}
\end{equation}
where $\erf()$ is the error function. Similarly, the time domain signal projected on the $\tau'$ axis is
\begin{equation}
    \begin{gathered}
        S_{proj,\omega_0}(\tau',T) = A(T) e^{-\frac{1}{4}[C(T)+1]\tau^{'2}\sigma^2}\int_{|\tau'|}^{\infty} e^{\frac{1}{4}([C(T)-1]t^{'2}\sigma^2-4\gamma t')} dt' \\
        =   \\
        \frac{\sqrt{\pi}}{\sigma \sqrt{1-C(T)}} e^{\frac{-\gamma^2}{\sigma^2 \left[C(T)-1\right]}-\frac{1}{4}\sigma^2 \tau'^{2}\left[1+C(T)\right]} \erfc{\left[\frac{\gamma}{\sigma \sqrt{1-C(T)}} + \frac{|\tau'|\sigma \sqrt{1-C(T)}}{2}\right]}.
    \end{gathered}
    \label{eqn:time-proj-tau-FFCF}
\end{equation}
Fourier transforms of eqns. \ref{eqn:time-proj-t-FFCF} and \ref{eqn:time-proj-tau-FFCF} cannot be performed analytically, therefore we will resort to doing them numerically to get the corresponding 2D frequency domain slices.
{Please see the Supplement for derivation of diagonal and cross-diagonal slices \cite{Supplement}.}

\begin{figure}[h!t]
    \centering
    \includegraphics[scale=0.4]{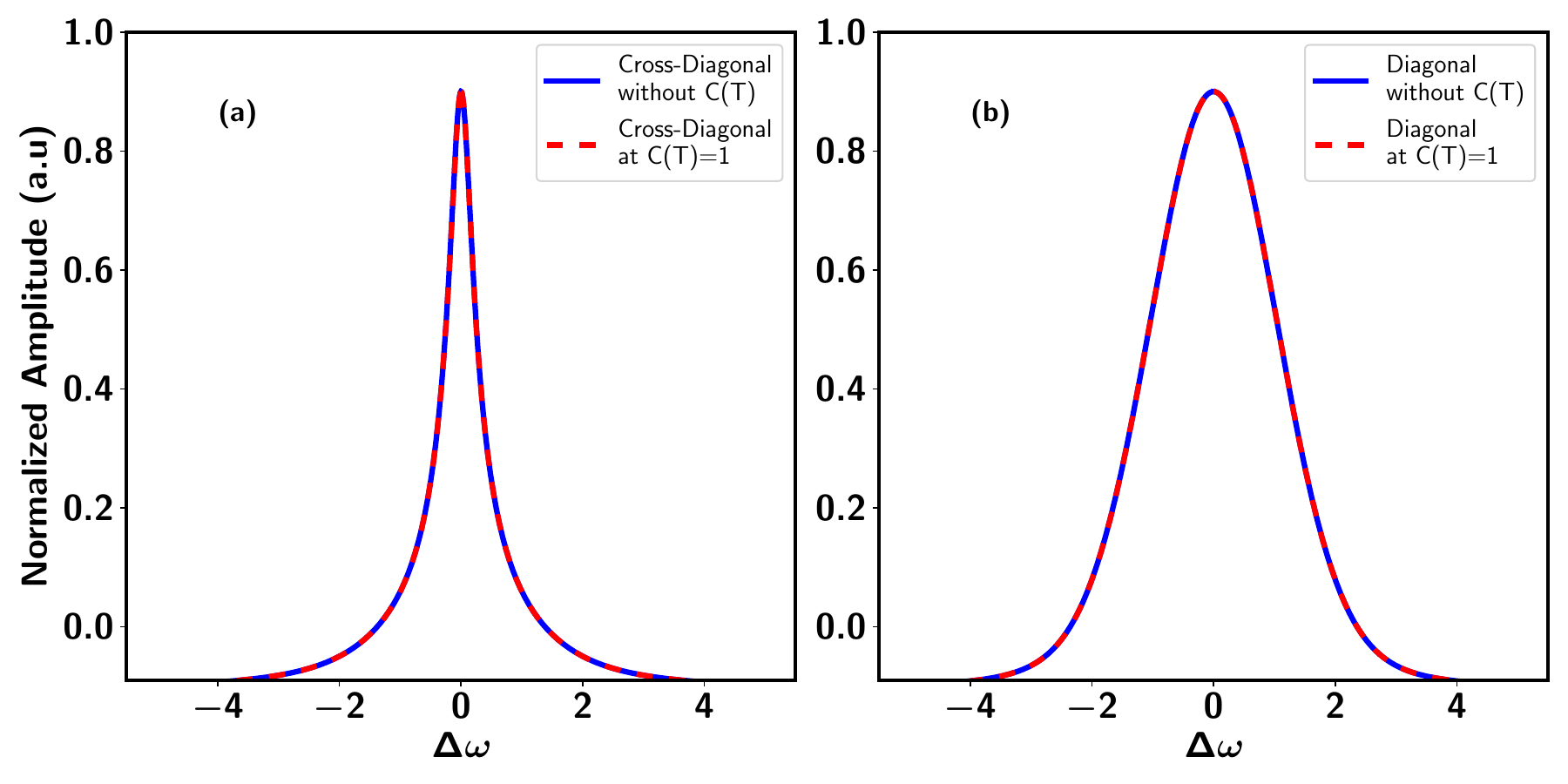}
    \caption{(a) Analyzing the cross-diagonal slice of both expressions with and without the Frequency-Frequency Correlation Function (FFCF) while setting C(T) = 1. (b) Comparison of diagonal slices.}
    \label{fig:comparison}
\end{figure}

In the limit that $C(T) = 1$, we obtain the same spectral slices as applying the PST to equation \ref{eqn:2Dtime-t'-tau'}, which can be Fourier transformed analytically. Performing the projection on to the $t'$ axis and Fourier transforming yields
\begin{equation}
    \begin{gathered}
        S_{proj,\omega_0}(\omega_{t'}) = \mathscr{F}\{S_{proj,\omega_0}(t')\}=\frac{e^{\frac{(-\gamma - i\omega_{t'})^2}{2\sigma^2}}\erfc{\left[\frac{\gamma-i\omega_{t'}}{\sqrt{2}\sigma}\right]}}{\sigma(\gamma-i\omega_{t'})}
    \end{gathered}
    \label{eqn:slice-omega_t'}
\end{equation}
where $\erfc$ is the complementary error function, and $\mathscr{F}$ denotes a Fourier Transform. Similarly, performing the projection on the $\tau'$ axis yields
\begin{equation}
    S_{proj,\omega_0}(\omega_{\tau'}) = \mathscr{F}\{S_{proj,\omega_0}(\tau')\}=\sqrt{\frac{2}{\pi \sigma^2}} e^{\omega_{\tau'}^{2} / 2\sigma^2} \circledast \frac{1}{\gamma^2 + \omega_{\tau'}^{2}}.
    \label{eqn:slice-omega_tau'}
\end{equation}
As a check, we compare Eq.~\ref{eqn:slice-omega_t'} and Eq.~\ref{eqn:slice-omega_tau'} with the numerical Fourier transforms of Eq.~\ref{eqn:time-proj-t-FFCF} and Eq.~\ref{eqn:time-proj-tau-FFCF} when $C(T)=1$, as shown in Fig.~\ref{fig:comparison}.

Both sets of Fourier transforms are equivalent as expected and we are now ready to examine the diagonal and cross-diagonal line shapes for other values of $C(T)$. As the system becomes less correlated, $C(T)$ approaches zero, the 2D spectra becomes more symmetric, and this expected trend is apparent in the diagonal and cross-diagonal line shapes shown in Fig. 2. 

\begin{figure}[h!]
    \centering
    \includegraphics[scale=0.78]{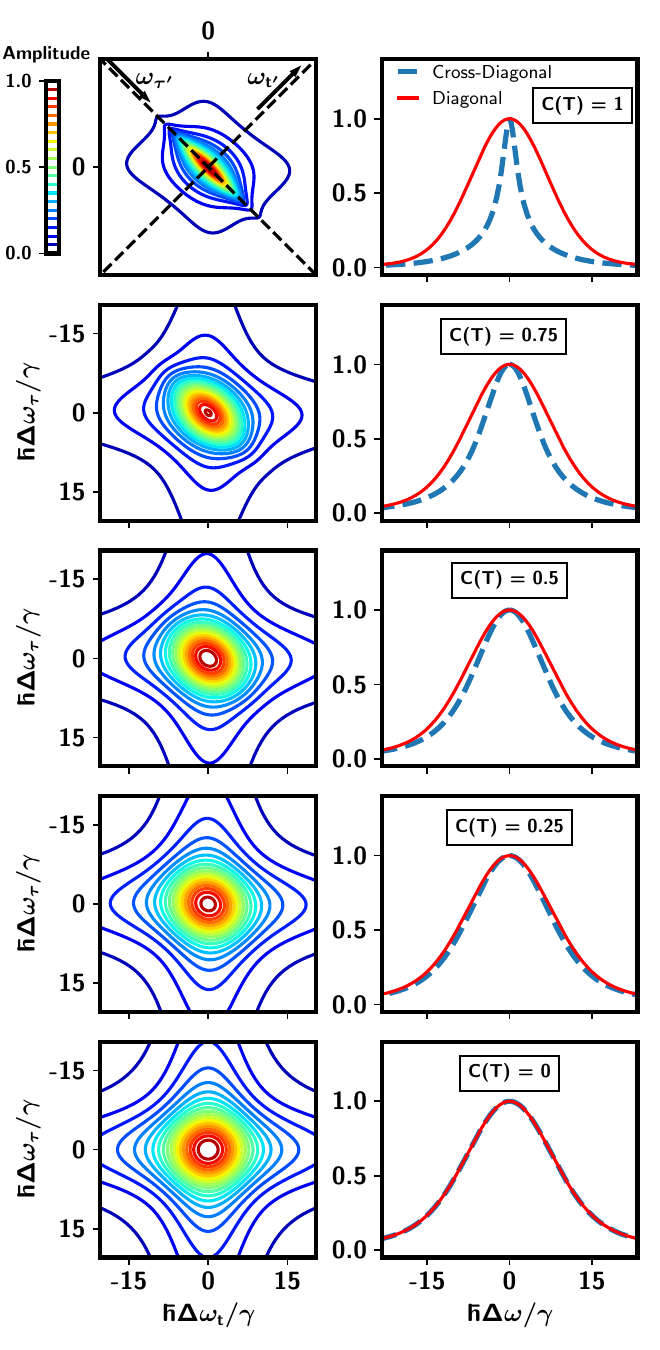}
    \caption{The sequence of 2D spectra (left column) and slices (right column) for a two-level system as $C(T)$ decreases from 1 to 0, arranged from top to bottom. The frequency detuning from the resonance frequency is normalized the the dephasing rate. The inhomogeneous distribution width is 5 times the homogeneous linewidth. For the two-dimensional spectra, the horizontal axis is the frequency corresponding to Fourier-transforming with respect to $t$ and the vertical axis with respect to $\tau$.}
    \label{fig:vary-C(T)}
\end{figure}

{The simulations shown in Fig.$~$2 were generated using the following parameters. The homogeneous and inhomogeneous broadening was set to $\gamma = 0.2$ meV and $\sigma = 2.0$ meV, respectively. For simplicity, the waiting-time-dependent amplitude, $A(T)$, was held at unity. The time domain signals were calculated on a $2048 \times 2048$ point grid, constructed from a time axis with a sampling of 50 fs, corresponding to a total range of approximately $\pm102.4$ ps for both the $\tau$ and $t$ axes.
}
\section{Fitting Procedure}

Physical parameters-the homogeneous dephasing rate $\gamma$, inhomogeneous broadening $\sigma$, the FFCF value $C(T)$, and $T$ dependent amplitude $A(T)$- are extracted from experimental data using a multidimensional nonlinear fitting procedure. The core method is to simultaneously fit the 1D diagonal and cross-diagonal slices from an experimental 2D spectrum to the theoretical model developed in the previous section. 
Let $M_{diag}(\omega_{\tau'};\gamma,\sigma,C(T),A(T))$ represent the theoretical lineshape for the diagonal slice, obtained by numerically Fourier transforming the time-domain projection in Eq. (6). Similarly, let $M_{crossdiag}(\omega_{t'};\gamma,\sigma,C(T),A(T))$ be the theoretical lineshape for the cross-diagonal slice, derived from Eq. (7).

To ensure a single, self-consistent solution, we construct a global cost function, $\Phi$, defined as the total sum of the squared residuals for both slices,

\begin{equation}
    \begin{gathered}
    \Phi(\omega_{\tau'},\omega_{t'};\gamma,\sigma,C(T)) = \\
    \sum_{i=0}^{n} (y_{i}^{diag} - M_{diag}(\omega_{\tau'};\gamma,\sigma,C(T),A(T)))^2 \\ +\sum_{i=0}^{n} (y_{i}^{crossdiag} - M_{crossdiag}(\omega_{t'};\gamma,\sigma,C(T),A(T)))^2
    \end{gathered}
    \label{eqn:approx-cost-mod}
\end{equation}
Here, $y_{i}^{diag}$ and $y_{i}^{crossdiag}$ are the data points of the experimental diagonal and cross-diagonal slices, respectively. The best fit values for the parameters are those that minimize this cost function. This minimization is performed using a standard iterative nonlinear least-squares algorithm \cite{1999Kelley_Book}, By fitting both slices at the same time, we leverage all available lineshape information and constrain the parameters in a physically meaningful way. This simultaneous approach is crucial because all four parameters influence both slices, ensuring a robust and self-consistent result that best describes the complete system dynamics.

\section{Fitting to Experimental Data}
To demonstrate our method and validate its necessity, we applied it to 2D coherent spectra of heavy-hole excitons in gallium arsenide (GaAs) quantum wells. The sample consists of four 20 nm-wide quantum wells grown by molecular beam epitaxy, held at a cryogenic temperature of 10 K. Co-circular polarization was used to isolate the heavy-hole exciton resonance. By performing rephasing scans at different population waiting times (T), we can track the evolution of the 2D lineshape and extract the FFCF. We analyze data at two key waiting times: a short delay (T = 0.2 ps) to validate the procedure, and a long delay (T = 60 ps) to demonstrate its importance. Further details of the experimental setup for acquiring the 2D spectra can be found in Ref.~\cite{2009Bristow_RSI}.

\begin{figure}[h!]
        \centering
        \includegraphics[width=1\linewidth]{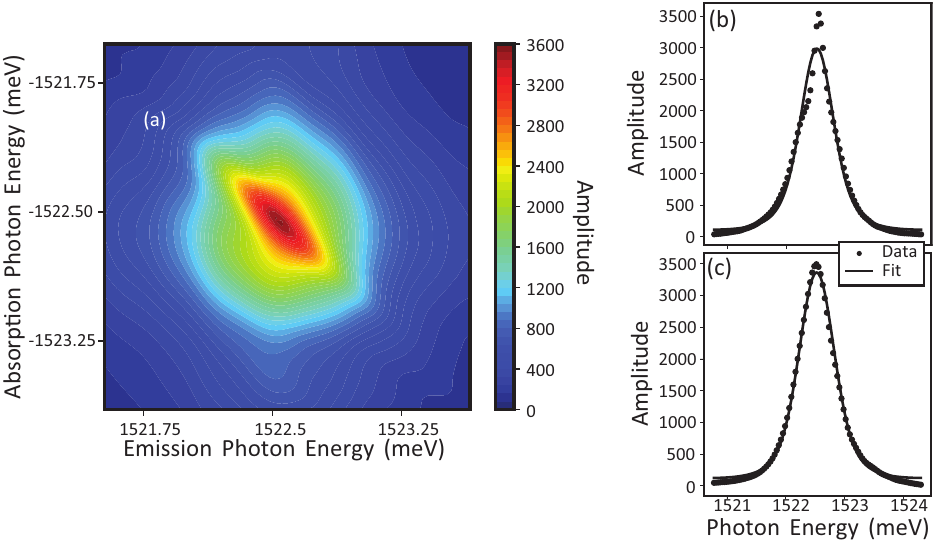}
        \vspace{-0.5cm}
        \caption{Fit of the experimental data for a short population waiting time of T = 0.2 ps. (a) 2D rephasing spectrum of the heavy-hole exciton. (b) Cross-diagonal and (c) diagonal slices (dots) with the fit results (solid lines). At this short delay, where $C(T) = 1$, the fitting procedure accurately extracts the lineshape parameters, and the inclusion of the FFCF term does not change the result.}
    \label{fig:main}
\end{figure}

{
First, we analyze the system at small $T$ where spectral diffusion is expected to be negligible. As shown in Fig.~3, the 2D spectrum at T = 0.2 ps exhibits distinct elongation along the diagonal, which is characteristic of a system with inhomogeneous broadening. In this regime, we expect the frequencies to be well correlated, meaning $C(T)=1$. Applying our simultaneous fitting procedure confirms this expectation. As detailed in Table 1, the fit yields $C(T)=1$. A simplified model that neglects the FFCF (equivalent to manually setting $C(T)=1$) produces an identical fit and the same linewidth parameters ($\gamma=0.217$ meV, $\sigma=0.154$ meV). We note that both models show deviation from the experimental data in a localized region near the peak in the cross-diagonal slice shown in Fig.~3(b). The fits match the experimental data very well in the diagonal slice shown in Fig. 3(c). We emphasize that both slices are fit simultaneously, thus the localized discrepancy does not have a significant effect and the fact that both models give similar parameters show that they are not being perturbed by the discrepancy. The discrepancy suggest that there are some additional physical processes not captured in model, however it could also be due to technical issues in the data collection.
}
\begin{figure}[h!]
    \centering
    
    \includegraphics[width=1\linewidth]{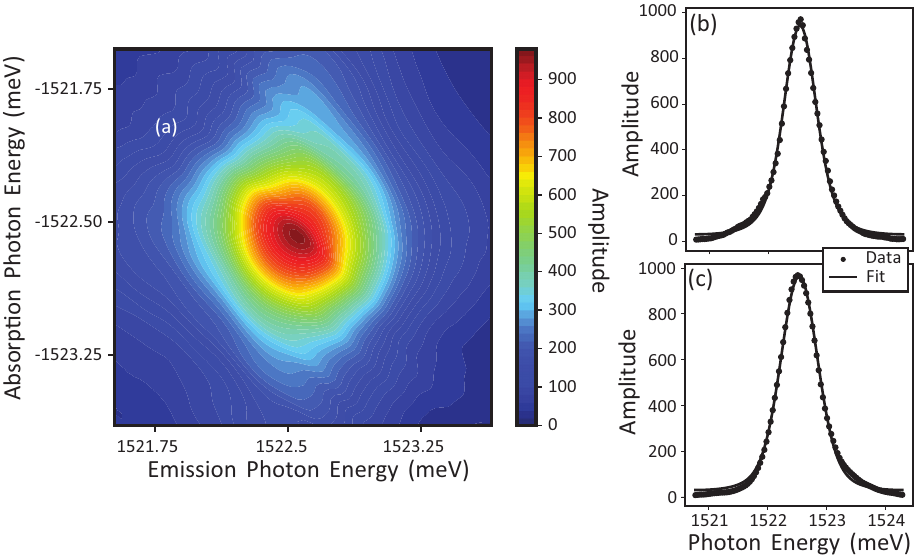}
        \vspace{-0.5cm}
        \caption{Fit of the experimental data at a long population waiting time of $T = 60~\mathrm{ps}$. (a) The 2D rephasing spectrum, showing a more symmetric lineshape due to spectral diffusion. (b, c) Fits (solid lines) to the cross-diagonal and diagonal data (dots). At this longer delay, including the FFCF is essential for a physically meaningful result, yielding $C(T) \approx 0.44$. Ignoring the FFCF leads to an unphysical determination of the system parameters.}
    \label{fig:enter-label}
\end{figure}

Next, we examine the system after a long delay of $T = 60~\mathrm{ps}$, allowing significant time for spectral diffusion to occur. The 2D spectrum, shown in Figure 4, is now visibly more circularly symmetric and less elongated, a clear qualitative signature of memory loss as excitons migrate within the quantum well.
{
We note that the cross diagonal slice shows a very weak asymmetry that is not present in the model. Such asymmetry can result from a breakdown of the strong redistribution approximation, however the asymmetry here is sufficiently weak such that the strong redistribution approximation is a reasonable assumption. 
}

The two fitting models diverge dramatically in terms of their estimates of the physical parameters, although the estimated goodness of fit parameter, $R^2$, is similar. The simplified model without the FFCF fails completely. In an attempt to describe the symmetric peak with an inherently asymmetric model, the fit returns a physically impossible result, a negative inhomogeneous broadening ($\sigma=-0.224$ meV). In stark contrast, our more complete model incorporating the FFCF provides an excellent fit to both the diagonal and cross-diagonal slices. This fit yields physically meaningful linewidths ($\gamma=0.187$ meV, $\sigma=0.27$ meV) and quantifies the remaining frequency correlations as $C(T)=0.445$. 

\begin{table}
    \centering
    \begin{tabular}{c||c|c|c|c|c}

\textbf{Method} & $T$ (ps) & $\gamma$ (meV) & $\sigma$ (meV) & $C(T)$ & $R^2$ \\
\hline\hline
\textbf{w/o FFCF} & 0.2 & 0.217 & 0.154 & -- & 0.9961 \\
\hline
\textbf{w/ FFCF} & 0.2 & 0.217 & 0.154 & 1.0 & 0.9961 \\
\hline
\textbf{w/o FFCF} & 60 & 0.307 & -0.224 & -- & 0.9975 \\
\hline
\textbf{w/ FFCF} & 60 & 0.187 & 0.27 & 0.445 & 0.9987 \\
\hline
\end{tabular}
    \caption{Parameters fitted at various delay times T using different fitting models, both with and without the Frequency-Frequency Correlation Function (FFCF). The $R^2$ scores are provided in the last column.}
    \label{tab:my_label}
\end{table}

The complete fitting results for both cases are summarized in Table 1. The comparison unequivocally demonstrates that while neglecting the FFCF is acceptable for very short times, it leads to erroneous and nonphysical conclusions when spectral diffusion is present. Our method, however, remains robust across all timescales, providing a direct and accurate measure of the underlying system dynamics. We note that the goodness of fit parameter, $R^2$, given in the last column of Table 1, is not significantly different for the fits at $T= 60~\mathrm{ps}$, showing that it is not a good indicator of the model without the FFCF, but rather the nonphysical negative inhomogeneous broadening signals its failure.

\section{Conclusion}

In conclusion, we have developed and experimentally validated a robust and streamlined method for quantifying spectral diffusion from two-dimensional coherent spectra. By directly incorporating the Frequency-Frequency Correlation Function (FFCF) into the Projection-Slice Theorem (PST) framework, our approach overcomes the limitations of both indirect geometric approximations and computationally expensive full-spectrum fitting routines. We demonstrated that this method accurately extracts the key system parameters—homogeneous and inhomogeneous linewidths alongside the FFCF—from experimental data on semiconductor quantum wells. Crucially, it succeeds in regimes where significant spectral diffusion is present, precisely where simpler models fail and produce nonphysical results.
{
The model does not invoke processes specific to our model system, semiconductor quantum wells, thus it is generically applicable to a broad range of systems.
}

{
We note that our derivations rely on the delta-pulse approximation, a common simplification in theoretical models of 2DCS. A more complete treatment considering finite-duration pulses, such as the analytical framework developed by Smallwood et al. \cite{Smallwood:17}, reveals that the frequency-domain spectrum is modified by a multiplicative "finite-pulse factor". This factor consists of Gaussian and error functions that not only limit the spectral bandwidth but also depend on the material's intrinsic resonance frequencies and decay rates \cite{Smallwood:17}. While these effects can introduce additional phase shifts and lineshape modifications , our impulsive-limit approach remains a robust and accurate method for lineshape analysis in the widely applicable regime where laser pulse durations are short compared to the system's dephasing dynamics.
}
The primary advantage of this technique is its combination of physical rigor and computational efficiency. By fitting 1D projections, it remains straightforward to implement while retaining a complete description of the system's dephasing dynamics. The simultaneous fitting procedure ensures a self-consistent and robust extraction of parameters without needing to decouple different dephasing mechanisms. Because the method makes no assumptions about the functional form of the FFCF decay, it is a versatile tool applicable to a wide variety of systems. This approach provides a new benchmark for lineshape analysis and offers a powerful addition to the toolkit for interpreting complex multidimensional coherent spectra.

\bibliography{bibliography}

\end{document}